\documentclass[12pt]{article}
\usepackage{epsf}

\newlength{\dinwidth}
\newlength{\dinmargin}
\def\lapproxeq{\lower .7ex\hbox{$\;\stackrel{\textstyle <}{\sim}\;$}}
\def\gapproxeq{\lower .7ex\hbox{$\;\stackrel{\textstyle >}{\sim}\;$}}

\def\be{\begin{equation}}
\def\ee{\end{equation}}
\def\bea{\begin{eqnarray}}
\def\eea{\end{eqnarray}}

\def\slashQ{/\!\!\! Q}
\def\slashk{/\!\!\! k}
\def\slashep{/\!\!\!\varepsilon}
\def\slashe{/\!\!\! e}
\def\slashp{/\!\!\! p}
\def\slashl{/\!\!\! l}
\begin{document}
\titlepage

\begin{flushright}
IPPP/04/58 \\
DCPT/04/116\\
17th September 2004 \\
\end{flushright}

\vspace*{4cm}

\begin{center}
{\Large \bf Inelastic $J/\psi$ and $\Upsilon$ hadroproduction }

\vspace*{1cm} \textsc{V.A.~Khoze$^{a,b}$, A.D. Martin$^a$, M.G. Ryskin$^{a,b}$ and W.J. Stirling$^{a,c}$} \\

\vspace*{0.5cm} $^a$ Department of Physics and Institute for
Particle Physics Phenomenology, \\
University of Durham, DH1 3LE, UK \\[0.5ex]
$^b$ Petersburg Nuclear Physics Institute, Gatchina,
St.~Petersburg, 188300, Russia \\[0.5ex]
$^c$ Department of Mathematical Sciences, 
University of Durham, DH1 3LE, UK \\%
\end{center}

\vspace*{1cm}

\begin{abstract}
We consider the  prompt hadroproduction of $J/\psi, ~\psi'$ and the
$\Upsilon (1S,2S,3S)$ states
caused by the fusion of a symmetric colour-octet state, $(gg)_{8s}$, and
an additional gluon.
 The cross sections are calculated in leading-order perturbative QCD.
 We find a considerable enhancement in comparison with previous perturbative QCD
 predictions.  Indeed, the resulting cross sections are found to be consistent with
the values measured at the Tevatron and RHIC, without the need to invoke
non-perturbative `colour-octet' type of contributions.
\vspace{1cm}
\end{abstract}

\section{Introduction}

\begin{figure}
\begin{center}
\centerline{\epsfxsize=0.6\textwidth\epsfbox{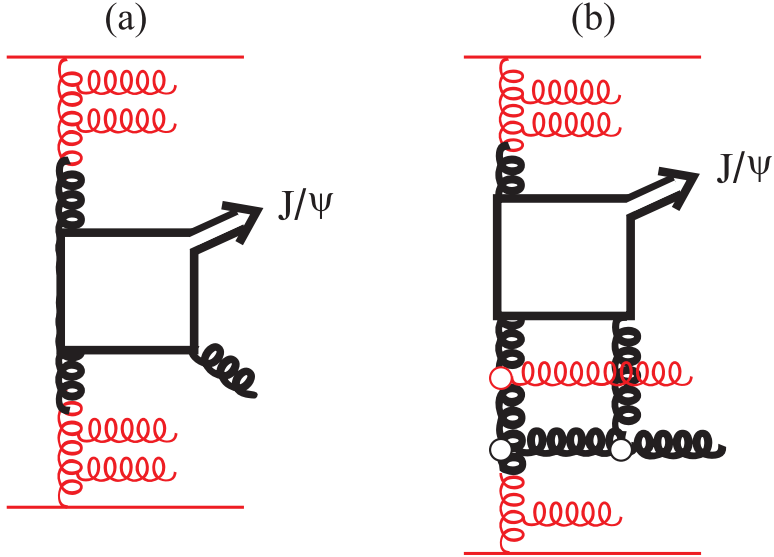}}
\caption{(a) The `bleaching' gluon subprocess used in the original `colour-singlet' perturbative QCD
estimates of prompt $J/\psi$ hadroproduction.  (b) The perturbative QCD mechanism
studied in this paper.   In each case the subprocess $gg \to J/\psi~g$ is shown
in bold.}
\label{fig:PQCD}
\end{center}
\end{figure}

It is not easy to describe the hadroproduction of $J/\psi$ mesons
within a perturbative QCD framework. The problem is that, due to the $J^P=1^-$
quantum numbers, it is not possible to directly form the colourless $J/\psi$ meson
by gluon-gluon fusion.
 The simplest possibility is to produce
a colour-octet quark-antiquark pair ($gg\to \bar cc$) and then to
emit an additional gluon, which carries away the colour, as shown in Fig.~\ref{fig:PQCD}a.
This is often referred to as the colour-singlet mechanism (CSM).  However the
corresponding cross section is suppressed by the small QCD coupling
$\alpha_s$, and by the additional phase space factor associated with the extra gluon
emission. As a result the LO QCD prediction \cite{BR} is found to be about  an order of
magnitude lower than the experimental yield of $J/\psi$ mesons.
% (see, for example \cite{Rev,Qia}
%and references therein for a more ). \\

An alternative and  more phenomenological approach is provided by the Colour Evaporation
Model (CEM) \cite{CEM}. Here the quarkonium production cross section is an (a priori unknown)
fraction of the $Q\bar Q$ heavy quark cross section integrated over the $m_{Q\bar Q}$ invariant
mass up to the threshold for producing a pair of the lightest heavy flavour mesons. There are
no constraints on the colour or spin of the $Q\bar Q$ pair, the transition from colour-octet $Q\bar Q $ 
to colour-singlet quarkonium is assumed to take place by the `evaporation' of soft gluons. The fraction
of $Q\bar Q$ pairs that materialise as a particular quarkonium state, $f_{c \bar c \to J/\psi}$ for example,
is assumed to be universal and is adjusted to give the best fit to existing data.\footnote{In practice, it 
will of course depend on the parameters and pdfs used in the calculation of the $Q\bar Q$ cross section.}
Despite its phenomenological success, the CEM has no firm theoretical foundation. In practice one would
not expect the evaporation of soft gluons to take place independently of the  particular collision
environment, and there is no reason why such soft interactions would not modify the quarkonium production 
properties, and in particular its collision energy dependence. This is precisely what happens in the
theoretically more rigorous formalism for quarkonium production that we introduce below.

The currently most popular and widely used description of quarkonium production is based on
a nonrelativistic QCD (NRQCD) effective field theory approach \cite{Oct}, which retains features of both
the CSM and CEM. Here $Q\bar Q$ pairs are produced via a hard partonic (short-distance) subprocess, in both
colour-singlet and colour-octet states, and non-perturbative universal matrix elements 
describe the (large-distance) transition
of the $Q\bar Q$ pair  into particular quarkonium states.\footnote{See, for example, \cite{Rev,Qia,Vogt}
and references therein, for a more detailed discussion of the situation, and of the history of the subject.}.
The cross section
of quarkonium $H$ production is written in the schematic form
\begin{equation}
d\sigma(H)\; =\; ~\sum_n d\hat\sigma(Q\bar Q[n])~\langle O^H[n] \rangle,
\label{OH}
\end{equation}
where $n$ denotes the set of colour and angular momentum quantum numbers of the
$Q\bar Q$ pair, and $\hat\sigma$ is the cross section of the $Q\bar Q$
pair production in a hard subprocess.
The non-perturbative transition from the $Q\bar Q$ state $n$ into the
 quarkonium state $H$ is described by a long-distance matrix element
 $\langle O^H[n] \rangle$.  These matrix elements are taken
 as parameters to be determined by fits to experimental data. In this way, it is
 possible to compensate the low value of the hard subprocess cross section,
 $d\hat\sigma(Q\bar Q[n])$, by the large fitted value of the matrix element
 $\langle O^H[n] \rangle$. Despite some phenomenological success, a detailed
proof of the factorisation formula (\ref{OH}) is lacking, and in particular
it is expected to break down at small values of the quarkonium transverse momentum.
Predictions for the {\it total} quarkonium cross section in the NRQCD approach
must therefore be treated with caution.

\begin{figure}
\begin{center}
\centerline{\epsfxsize=0.8\textwidth\epsfbox{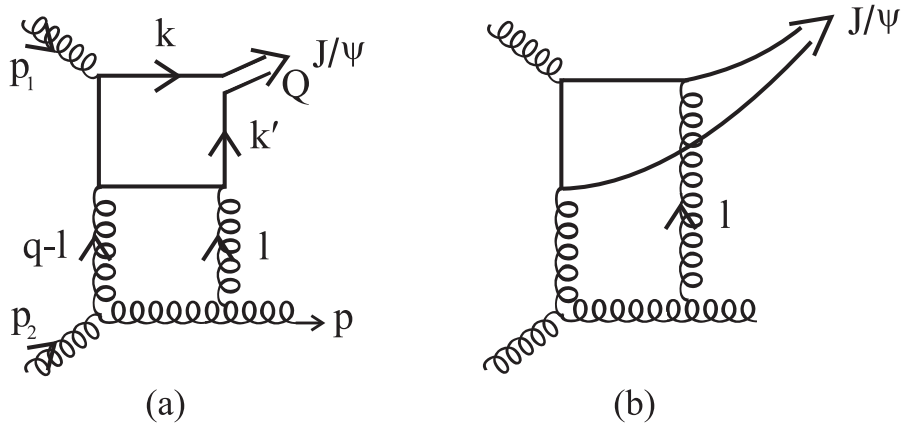}}
\caption{Lowest-order perturbative QCD diagrams for $J/\psi$ hadroproduction
via gluon-gluon fusion with an additional gluon.}
\label{fig:H}
\end{center}
\end{figure}

Since the mass of the $J/\psi$ meson is not particularly small, it
 would be desirable to be able to describe the $Q\bar Q\to H$ transition within
 a perturbative QCD framework. In other words, we would like to
 consider explicitly an extra gluon exchange (which was {\it hidden}
 in the value of the non-perturbative matrix element $\langle O^H[n] \rangle$ in the case
 of the colour-octet mechanism, i.e. where the $J/\psi$ meson is formed from the 
colour-octet $Q\bar Q$ pair after some non-perturbative interaction described by the
$\langle O^H[n] \rangle$ matrix element. The corresponding lowest order in $\alpha_s$ 
 diagrams are shown in Fig.~\ref{fig:H}. In comparison with the $gg\to Q\bar Q$
 amplitude, the contributions of Fig.~\ref{fig:H} contain an extra loop
 factor with a small coupling $\alpha_s$. Here we investigate the
 possibility that this suppression is
 compensated by the large number of graphs
 where the additional (third) gluon, needed to form the $J/\psi$, is absorbed
 by different parton-spectators.  As viewed from the collinear approximation,
 the amplitude shown in Fig.~\ref{fig:PQCD}b, and Fig.~\ref{fig:H}, corresponds to the
 next-to-next-to-leading order (NNLO) contribution to the cross section (as
 compared with LO $gg \to \chi_c$ production).
 However the second $t$-channel gluon may be absorbed by {\it any} parton spectator;
 that is the amplitude is enhanced by the parton multiplicity, $n \propto {\rm log} s$.
Therefore it may be considered as the LO amplitude in the BFKL approach.
The aim of this paper is to evaluate the numerical size of this enhanced `NNLO'
contribution, and to see if it can remove the large discrepancy between the
perturbative QCD prediction and the inelastic $J/\psi$ hadroproduction data.

 In Section 2 we calculate the amplitude of the process shown in Fig.~\ref{fig:H}
 using the non-relativistic $Q\bar Q\to J/\psi$ vertex that was proposed in
 Ref.~\cite{BJ}. Due to the non-relativistic wave function of
 the $J/\psi$, there is practically no integration over the quark loop in
 Figs.~\ref{fig:H}a,b.   Indeed, the $J/\psi$ vertex (i.e. the $J/\psi$ wave function integrated
 over the relative momenta of the charm quarks), together with the two
 nearest $c$-quark propagators is
 \be
 g (\slashk +m)\gamma_\nu,
 \ee
 where the index $\nu$ corresponds to the $J/\psi$ polarization vector; and
 $k_\mu=Q_\mu/2$ and $m$ are the 4-momentum and the mass of the $c$-quark
 ($Q_\mu$ is the momentum of the $J/\psi$).  The constant $g$ may be
 expressed in terms of the electronic width $\Gamma^J_{ee}$ of the
 $J/\psi\to e^+e^-$ decay
 \begin{equation}
g^2\;=\; \frac{3\Gamma^J_{ee}M}{64\pi\alpha^2},
\label{gee}
\end{equation}
where $M=2m$ is the mass of the $J/\psi$ meson, and the electromagnetic coupling
 $\alpha=1/137$.

In Section 3 we compute the total cross section of {\it prompt}
inelastic $J/\psi$ hadroproduction at collider energies.  That is,
in that section, we neglect the additional
$J/\psi$ yield coming from $b$-quark or $\chi_c$ decays. We find that the LO
result agrees with the Tevatron data rather well. The transverse
momentum distributions of prompt $J/\psi$, $\psi'$
and the $\Upsilon(1S,2S,3S)$ production are
presented in Sections 4 and 5, together with a brief discussion of the $J/\psi$ polarization.
Alternative possibilities to produce the
 $J/\psi$ are considered in Section 6.  One such mechanism is to create a colour-octet
$c\bar c$ pair, which then transforms to a colour singlet by
 rescattering via gluon exchange. Another is the associative production
 of a $J/\psi$ and $c\bar c$ pair. The yield from this latter possibility is small. However
  the first mechanism, where two gluons in a symmetric colour-octet
$t$-channel state, $(gg)_{8s}$, belong to two different Pomerons, may
 dominate at asymptotically large energies. The energy and rapidity
 dependence of $J/\psi$ (and $\Upsilon$) production is given in Section 7. Section 8 contains our
 conclusions.

\section{The lowest-order amplitude}
We compute the matrix element of the hard subprocess from the diagrams
of Fig.~\ref{fig:H} in the LO collinear approximation. Thus incoming
 particles, with momenta $p_1$ and $p_2$, are taken to be
on-mass-shell, transversely-polarized gluons.
The calculation of the amplitude shown in Fig.~\ref{fig:H}a is similar
to the computation of the amplitude for diffractive $J/\psi$
photoproduction \cite{RJ}. Due to the non-relativistic nature of
the $J/\psi$ wave function, the difference between the quark momenta
$k$ and $k'$ is very small, that is $|k-k'| \ll m$. Therefore, following Ref.~\cite{BJ},
 we may take
$k=k'=Q/2$, and include the integration over the quark loop momentum
in the $c\bar c\to J/\psi$ coupling $g$, which is normalized to the
width of $J/\psi\to e^+e^-$ decay, see (\ref{gee}).

We expect that the main contribution will come from the region where
the rapidity difference between the final gluon $p$ and $J/\psi$ meson
is rather large; that is $s=(p_1+p_2)^2 \gg M^2$. In this limit, the
amplitude $A^a$, corresponding to the diagram Fig.~1a, is
\be
{\rm Im} A^a\;=\; \frac{N_c}8d^{abc}\int dl^2_t~g(4\pi\alpha_s)^{5/2}
~\frac{{\rm Tr}[\slashe(\slashQ /2+m)~\slashep\slashp_2(-\slashQ /2-\slashl+m)
\slashp_2(\slashQ /2-\slashp_1+m)]}
{2\pi s[(Q/2-p_1)^2-m^2][l^2-\lambda_g^2][(q+l)^2-\lambda_g^2]},
\label{Aa}
\ee
where $\varepsilon$ and $e$ are respectively the polarization vectors of the $J/\psi$ and the gluon
with momentum $p_1$, and $q=p_1-Q$ is the momentum transferred
through the pair of $t$-channel gluons.
$N_cd^{abc}/8$ is the colour factor, where the indices $a,b,c$ are the colours of the
two incoming gluons and the
final gluon. Bearing in mind possible
confinement effects (and to avoid the logarithmic infrared
singularity as $q\to 0$)
we introduce a cutoff (or effective gluon mass) $\lambda_g$
 in the denominator of (\ref{Aa}).
Since the two $t$-channel gluons
are in a symmetric colour-octet state, the amplitude $A$ has positive
signature. Therefore it has a small real part, ${\rm Re}A \ll {\rm Im}A$.

To obtain the whole amplitude, $A$, we have to add the contribution of
the diagram shown in Fig.~\ref{fig:H}b, which is of the form
\be
{\rm Im}A^b\;=\; -\frac{N_c}8d^{abc}\int dl^2_t~g(4\pi\alpha_s)^{5/2}
~\frac{{\rm Tr}[\slashe(\slashQ /2-\slashl+m)\slashp_2(-\slashQ /2+m)
~\slashep
\slashp_2(\slashQ /2-\slashl-\slashp_1+m)]}
{2\pi s[(Q/2-l-p_1)^2-m^2][l^2-\lambda_g^2][(q+l)^2-\lambda_g^2]},
\label{Ab}
\ee
where the minus sign reflects the `negative' colour charge of the antiquark.
We also have to account for the contributions where the gluon $p_1$ couples to the
antiquark, and not to the quark. That is, we must include the graphs with the opposite
direction of the `arrows' in the quark loop. Thus the differential
cross section is
\begin{equation}
\frac{d\hat\sigma}{dq^2_t}\;=\;
\frac{|A|^2}{16\pi s^2}\; ,
\label{dsdt}
\end{equation}
where the total amplitude $A=2(A^a+A^b)$.

For transverse (with respect to the $p_{1\mu},p_{2\mu}$
plane) $J/\psi$ inelastic production the trace
Tr$[...]=s^2m(e\cdot\varepsilon)$, while
Tr$[...]=s^2m(e\cdot Q_t)/M$ when the vector $\varepsilon$
 corresponds to a longitudinally polarized $J/\psi$ meson. Thus,
 after averaging over the incoming gluon transverse polarizations,
 $e^\perp$, the ratio of longitudinal ($\hat\sigma^L$) and transverse
 ($\hat\sigma^T$) cross sections becomes
\begin{equation}
\frac{d\hat\sigma^L/dq^2_t}{d\hat\sigma^T/dq^2_t}\;=\;
 \frac{|Q^2_t|}{2M^2}\, .
\label{dsLT}
\end{equation}

\section{The prompt $J/\psi$ yield}
In the LO collinear approximation, the cross section of inelastic prompt
$J/\psi$ production is of the form
 \begin{equation}
 \frac{d\sigma}{dydQ^2_t}=\int\frac{dx_2}{x_2}x_1g(x_1)x_2g(x_2)
 \frac{d\hat\sigma(s,q^2_t)}{dq^2_t}\; ,
\label{sigt}
\end{equation}
where $y$ is the centre-of-mass  rapidity of the $J/\psi$ meson,
 $x_1=s/(x_2S)$,
%  $x_{1,2}=(M_\perp/\sqrt S)\exp{(\pm y)}$,
% $M_\perp=\sqrt{M^2+|Q^2_t|}$,
 $S$ is the initial hadron-hadron energy
 squared and
 $x_ig(x_i)$
 are the densities of gluons in the incoming hadrons ($i=1,2$).
For a fixed rapidity of the $J/\psi$ meson, the integral over $x_2$
is equivalent to the integration over the mass $\sqrt s$ of the $(J/\psi+g)$ system,
where $s=(p_1+p_2)^2$.

 The hard subprocess cross section $\hat\sigma$ includes the
 contribution of the diagrams Fig.~\ref{fig:H}, where the single gluon $p_1$ comes
 from the beam side, plus the `inverse' contribution, in which the
 single gluon comes from the target. There is no interference between
the original (Fig.~\ref{fig:H}) and  the `inverse' amplitudes, due to the different
 colour structure of  single and double gluon exchange\footnote{The
 interference would correspond to {\it odderon},
 instead of Pomeron, exchange in the diagram for the cross
 section.  It is known, both experimentally and theoretically, that the
 odderon-nucleon coupling is small.}. Note that single and double gluon exchange
 correspond, respectively, to antisymmetric and symmetric colour octets.

 In the small $x$ region, and at relatively low scales, the gluon
 distribution behaves like $xg(x)\propto x^{-\lambda}$, where the power $\lambda\sim
 0.2$, while the `hard' subprocess cross section $\hat\sigma$ does not depend on
 $s$ for $s \gg M^2$.  Thus the integral over $x_2$ takes the form $\int
 dx_2/x_2^{1+\lambda}$. The main contribution comes from the lowest
 values of $x_2\simeq M_\perp {\rm e}^{-y}/\sqrt S$, which correspond to
$x_1\simeq M_\perp {\rm e}^{y}/\sqrt S$, where 
% $M_\perp=\sqrt(M^2+|Q^2_t|)$.
$M_\perp=(M^2+|Q^2_t|)^{\frac{1}{2}}$.
  However the essential
 interval of integration available at collider
 energies ($\Delta\ln{x_2}\approx 1/\lambda\sim 5$) is quite large.
 This large integration interval partly compensates for the small loop
 ($\alpha_s$) factor in the $g(gg)\to J/\psi + g$ amplitude
 obtained from the diagrams of Fig.~\ref{fig:H}.

\begin{figure}
\begin{center}
\centerline{\epsfxsize=0.5\textwidth\epsfbox{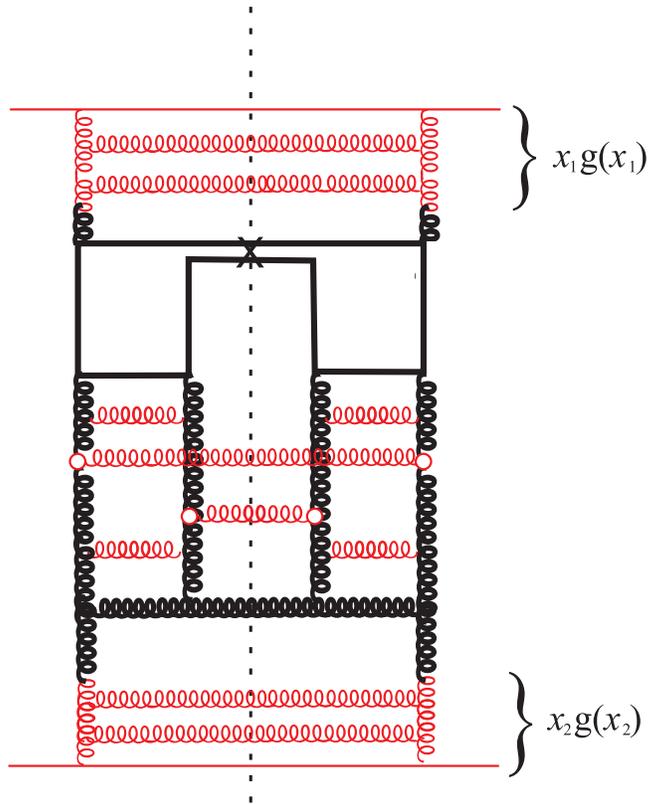}}
\caption{A contribution to inclusive prompt $J/\psi$ production accompanied by gluon emissions from a
$t$-channel gluon-pair in a colour-symmetric octet.  Again, the hard subprocess is shown
by bold particle lines.}
\label{fig:S}
\end{center}
\end{figure}

 Moreover for an inelastic process we have to allow for the
 emission of additional (secondary) $s$-channel gluons from the symmetric
 octet $(gg)$ (see Fig.~\ref{fig:S}). This leads to a power growth of the `hard'
 subprocess cross section, $\hat\sigma(s)$, as the function of
 subenergy $\sqrt s$. The power
 behaviour of $\hat\sigma(s)\sim s^\delta$ is driven by the intercept
 of the amplitude with the four $t$-channel gluons (the so-called quarteton).
 Within the leading logarithm approximation ($\sum_n C_n(\alpha_s\ln s)^n$)
this intercept was evaluated in \cite{LV}. It was
 shown that numerically the value of $\delta$ is, to within 10\%, equal
  to that for the usual BFKL Pomeron (two-gluon)
 exchange.
As a consequence we have to extend the $dx_2/x_2$ integration over the
 whole kinematically available rapidity interval $\Delta y$. Then the
 prompt $J/\psi$ cross section becomes

 \begin{equation}
 \frac{d\sigma}{dydQ^2_t}=x_1g(x_1)x_2g(x_2)\Delta y
 \frac{d\hat\sigma(s,q^2_t)}{dq^2_t}\; ,
\label{sigf}
\end{equation}
with
  $x_{1,2}=(M_\perp/\sqrt S)\exp{(\pm y)}$,
 and $\Delta y=\ln(x_{\rm max}^2 S/M^2_\perp)$. We take the same scale, $M_\perp$, for both gluons; 
different scales are
equivalent to a NLO correction to the LO formula. We introduce the factor $x_{\rm max} =0.3$
 to exclude the contribution from the graphs in which the third gluon
 couples to partons with large $x_{1,2}>0.3$ in the proton fragmentation
 regions, that is to allow for the fact that parton densities at large $x$ are kinematically
 suppressed.\footnote{The variation of $x_{\rm max}$ induces a NLL correction; that is, it is
equivalent to an $\alpha_s$ term (without
a $\ln(1/x)$ factor) in the BFKL amplitude.
Indeed, in the BFKL expansion, the contribution of Fig.~\ref{fig:PQCD}b to the cross section
behaves as an $\alpha_s \Delta y$ term as compared to that of  Fig.~\ref{fig:PQCD}a.
Thus the variation of $x_{\rm max}$ corresponds to an $\alpha_s$
term without a $\ln(S/M^2)$ factor.
  This does not mean, of course, that the variation of $x_{\rm max}$ reproduces
the {\it whole}
NLL contribution to the BFKL intercept.}

 The formulae in Eqs.~(\ref{sigf}) and (\ref{dsdt}), together with
 the amplitudes of (\ref{Aa},\ref{Ab}), enable the
 inelastic $J/\psi$ cross section to be predicted.
In this way, we find that the cross section at the Tevatron energy $\sqrt{S}=1.96$ TeV is\footnote{We
use the LO MRST2001 \cite{MRST} gluon distribution at scale $\mu=M_\perp/2$, with the corresponding
LO (one loop) QCD coupling $\alpha_s$ with $\Lambda^{(4)}_{\rm QCD}=220$~MeV.}
\be
\sigma(|y|<0.6)~\simeq ~2.7~\mu{\rm b}~~~~~~~~~~~{\rm (pQCD ~estimate)}
\label{sig}
\ee
in the central rapidity interval $|y|<0.6$ (integrated over the
transverse momentum $Q_t$).  Here we have taken\footnote{The effective
gluon mass $\lambda_g$, which occurs in the amplitudes, was first estimated in
Ref.~\cite{PP}, where it was introduced to describe the photon spectra in $J/\psi \to \gamma gg$
decay.  A recent evaluation, together with a collection of previous determinations, is
presented in Table 15 of Ref.~\cite{field}.  Based on this Table we choose
$\lambda_g=0.8$ GeV, with a possible uncertainity covered by the interval $\lambda_g=0.5~-~1$ GeV.}
 $\lambda_g=0.8$ GeV.
This prediction, (\ref{sig}), is to be compared with the lastest experimental measurement
\cite{CDF}
\be
\sigma(|y|<0.6)~= ~4.1~^{+0.6}_{-0.5}~\mu{\rm b}~~~~~~~~~{\rm (CDF ~experiment)}
\label{expt}
\ee

The uncertainties of the prediction, (\ref{sig}), are as follows.

(i) The choice of $\lambda_g$.   If, instead of 0.8 GeV, we were to take
$\lambda_g$ equal to 0.5 or 1 GeV then $\sigma(|y|<0.6)$ becomes 4.0 or
2.0 $\mu$b respectively.

(ii) The choice of the factorization and renormalization scales. For different
scales $\mu=\mu_R=\mu_F$ of $M_\perp/2,~M_\perp$ and $2M_\perp$,  we obtain
 $\sigma(|y|<0.6)~=~2.7, ~2.3$ and $1.5~\mu$b, respectively.
 
(iii) An unknown K-factor to account for NLO and higher pQCD corrections.

(iv) The uncertainty in the incoming gluon distribution, which is not
well constrained at low $x$ and rather low scales.

(v) The choice of the cut-off $x_{\rm max}$. The variation of the value of
$x_{\rm max}$ plays the role of NLL corrections in the BFKL approach.\\

Taking all these into account, the expected accuracy of the prediction is about a factor
of 2$-$3 in either direction or even worse.

\section{Transverse momentum distribution}

Unfortunately we cannot use the amplitudes of Eqs.~(\ref{Aa},\ref{Ab})
directly to calculate the $Q_t$ distribution of the produced $J/\psi$
mesons. In the case of diffractive $J/\psi$ photoproduction
it is known that the interference between the two lowest-order
diagrams, Fig.~\ref{fig:H}a and Fig.~\ref{fig:H}b, leads to a dip in the
$Q_t$ distribution.  Indeed, the differential
cross section goes to zero at $Q_t=M$ \cite{FR,BFL}. However this dip disappears
after one includes the leading logarithmic ($\alpha_s\ln S$)
correction \cite{FR,BFL}. The problem is that we cannot resum the
analogous corrections in our case, since the corresponding
`quarteton' eigenfunctions (coming from 4 gluons in the $t$-channel) are not yet known.

\begin{figure}
\begin{center}
\centerline{\epsfxsize=0.8\textwidth\epsfbox{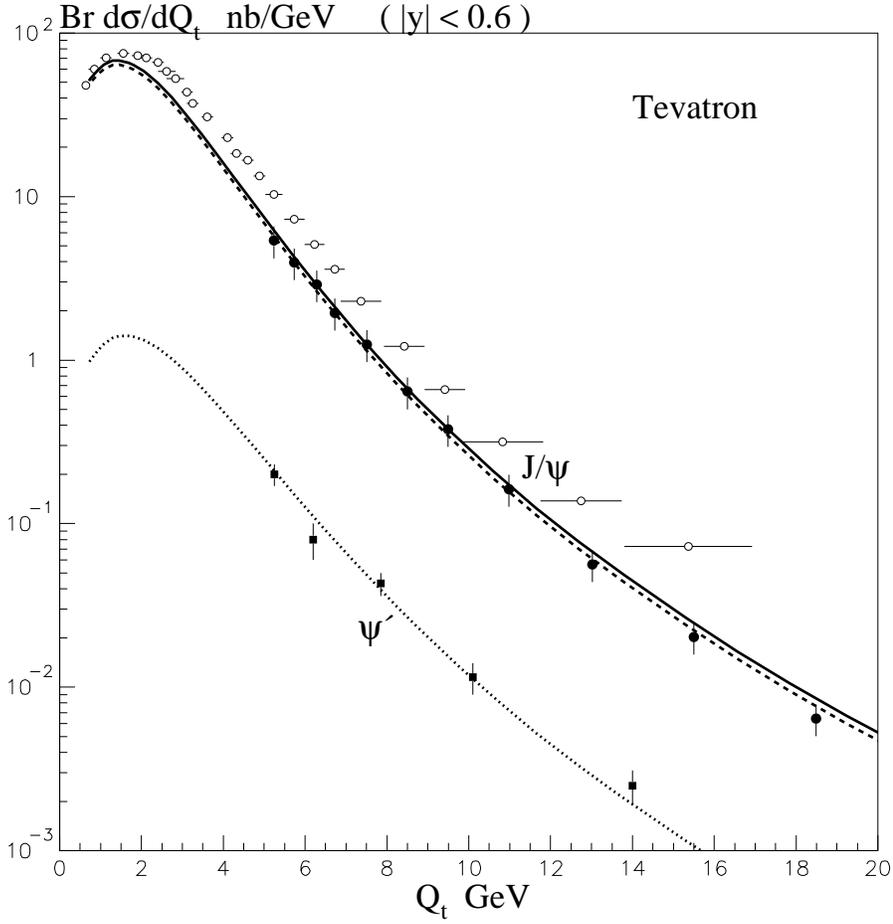}}
\caption{The transverse momentum ($Q_t$) distributions of inelastic $J/\psi$ and $\psi'$ production.
The data are from Refs.~\cite{CDF,psiprime}.  The upper and lower data sets for
the $Q_t$ distribution of the $J/\psi$ correspond to the total (at $\sqrt S$ = 1.96 TeV) and
prompt (at $\sqrt S$ = 1.8 TeV) $J/\psi$ yields respectively; recall that our QCD prediction is
for prompt production only.}
\label{fig:qt}
\end{center}
\end{figure}

Therefore we consider a very simple parametrization
\begin{equation}
\label{Qt}
d\hat\sigma/dQ^2_t~~\propto~~ g^2\alpha_s(M_\perp)^5 {\rm log}(x_{\rm max}^2 S/M^2_\perp)/M^6_\perp
\end{equation}
motivated by dimensional counting, which accounts for the
dimension of $g^2\sim \Gamma M$.  Again we take $x_{\rm max}=0.3$.
The distribution (\ref{Qt}) is normalised by equating its $Q_t^2$ integral to that
of (\ref{sigf}).
In this way the effective gluon mass, $\lambda_g$, enters the calculation.
The result, shown in Fig.~\ref{fig:qt}, is
in reasonable agreement with the Tevatron data \cite{CDF,psiprime}.

In comparison with the colour octet model, where the $J/\psi$ is
created in the fragmentation of a gluon jet and the expected
distribution is $d\sigma/dQ^2_t\propto 1/Q^4_t$,
in the case of the Fig.~\ref{fig:S} subprocess the  $Q_t$ distribution (\ref{Qt})
at large $Q_t$ is steeper, $d\sigma/dQ^2_t\propto 1/M^6_\perp$.
This is similar to the distribution in the colour-singlet model~\cite{BR}.
However in contrast with the colour-singlet mechanism (Fig.~\ref{fig:PQCD}a)
here the hadronic transverse energy flow, which compensates the $Q_t$ of
the $J/\psi$ meson, is distributed over a larger rapidity interval
in the form of  a larger number of gluonic minijets (see
Fig.~\ref{fig:S}).

 According to (\ref{dsLT}) we expect the $J/\psi$ mesons to be
 transversely polarized at small $Q_t$, and longitudinally polarized at
 large $Q_t$, that is $Q_t \gg M$.\footnote{At next-to-leading order the polarization of
the $J/\psi$ may also be affected by the contribution of the
longitudinally polarised incoming gluons ($p_1$ in Fig.~\ref{fig:H}a),
see for example Ref.~\cite{NNN}.}
 At present the data are only available in the
 interval of $Q_t\sim 4~-~20$ GeV, where the observed $J/\psi$ is
 approximately unpolarized. The parameter $\alpha<0.3$ of Ref.~\cite{polar}
corresponds to a small transverse polarization at lower $Q_t$ values.
However, as $Q_t$ increases $\alpha$ changes sign and for $Q_t>15$ GeV clearly indicates
 longitudinal polarization of the $J/\psi$.
   This is qualitatively consistent with our expectations. In contrast, 
colour-octet models of prompt $J/\psi$ production lead to transverse polarization 
at large $Q_t$ \cite{polthy}.

 \section{Prompt $\psi'(2S)$ and $\Upsilon$ production}

\begin{figure}
\begin{center}
\centerline{\epsfxsize=0.8\textwidth\epsfbox{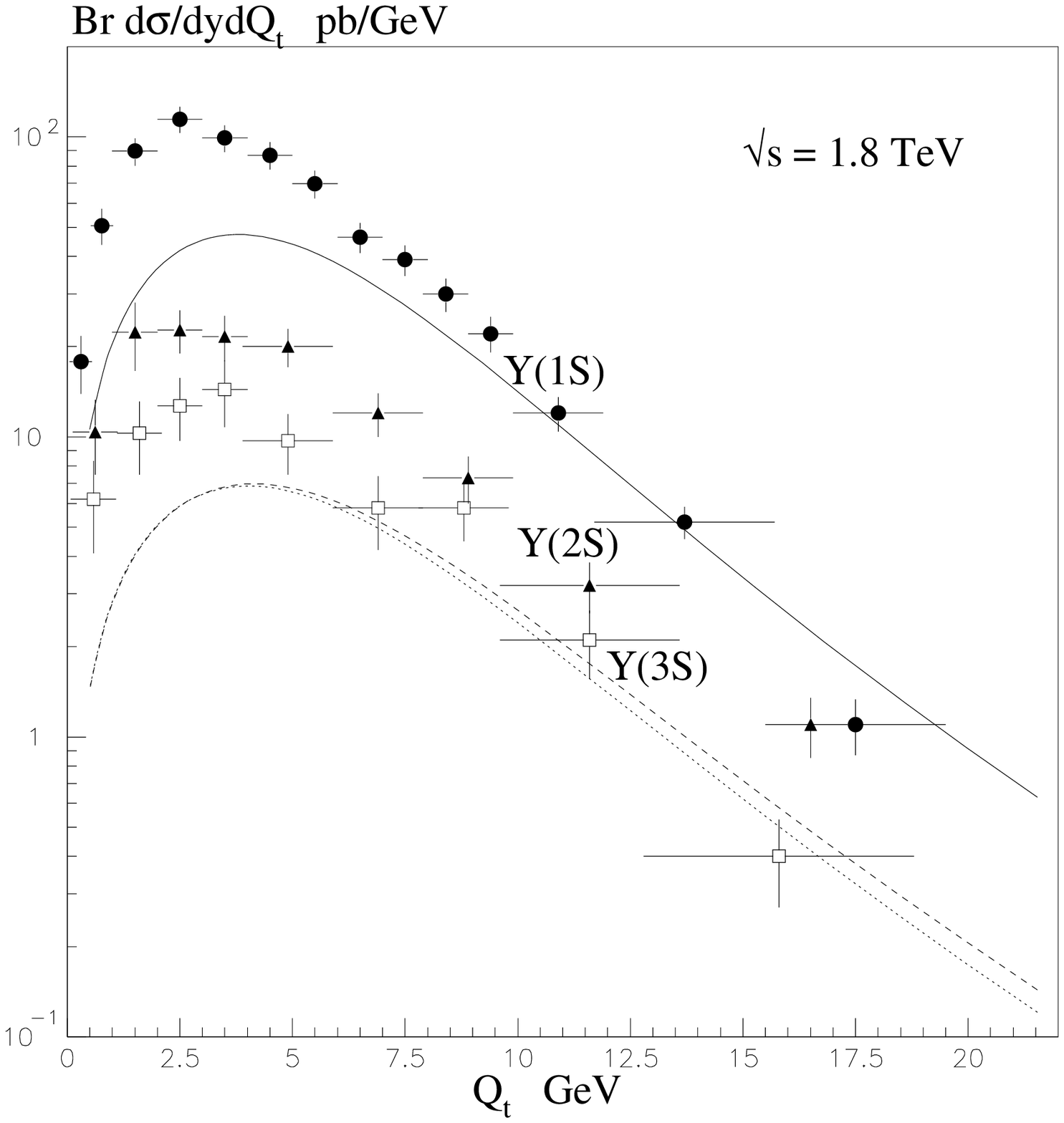}}
\caption{The $Q_t$ distributions of the inelastic production of the $\Upsilon$
states, compared with Tevatron data \cite{up1}. }
\label{fig:up}
\end{center}
\end{figure}

 The above formalism can be applied to the production of other quarkonium states by 
simply changing the mass and width of the $J^P=1^-$ heavy $Q\bar Q$
 resonance.  Thus, without any free parameters, we can predict the cross section
 for inclusive prompt quarkonium production.  The results for $\psi'$ and for
 the upsilon states are compared with the Tevatron data \cite{psiprime,up1} in
Fig.~\ref{fig:qt} and Fig.~\ref{fig:up} respectively.  We use the effective
gluon mass $\lambda_g$ and the scale that were appropriate for the
description of the $J/\psi$ data.  In this way we remove a large part of
the uncertainty in the prediction for $\psi'$.  The good agreement with
the $\psi'$ data should therefore be regarded as support for our
perturbative QCD approach.

The comparison of the prompt $\Upsilon$ predictions with the Tevatron data,
shown in Fig.~\ref{fig:up}, is complicated, however, since only half of the total $\Upsilon(1S)$
yield arises from prompt production \cite{up2}.  Moreover, the $\chi_b$ states
have a large branching fraction of radiative decays to $\Upsilon(2S)$; about
twice as large as those to $\Upsilon(1S)$.  Bearing in mind these complications,
and the uncertainties in the predictions, the agreement with the data is better
than may have been expected.

The predicted cross sections for prompt $J/\psi,~\psi';~\Upsilon(1S,2S,3S)$ central production
at the Tevatron energy, $\sqrt S$ = 1.96~TeV, are
\be
\left. d\sigma /dy\right\vert_{y=0}~~~=~~~2.2,~0.6~\mu{\rm b};~~~40,~12,~9~{\rm nb},~~~~~~~~~~~~~{\rm (Tevatron)}
\ee
respectively; and correspondingly at the LHC energy, $\sqrt S$ = 14~TeV,
\be
\left. d\sigma /dy\right\vert_{y=0}~~~=~~~8.1,~2.5~\mu{\rm b};~~~310,~100,~80~{\rm nb}.~~~~~~~~~~~~~{\rm (LHC)}
\ee

 \section{Other production mechanisms}

 \subsection{Uncorrelated gluon-gluon pairs}
\begin{figure}
\begin{center}
\centerline{\epsfxsize=0.8\textwidth\epsfbox{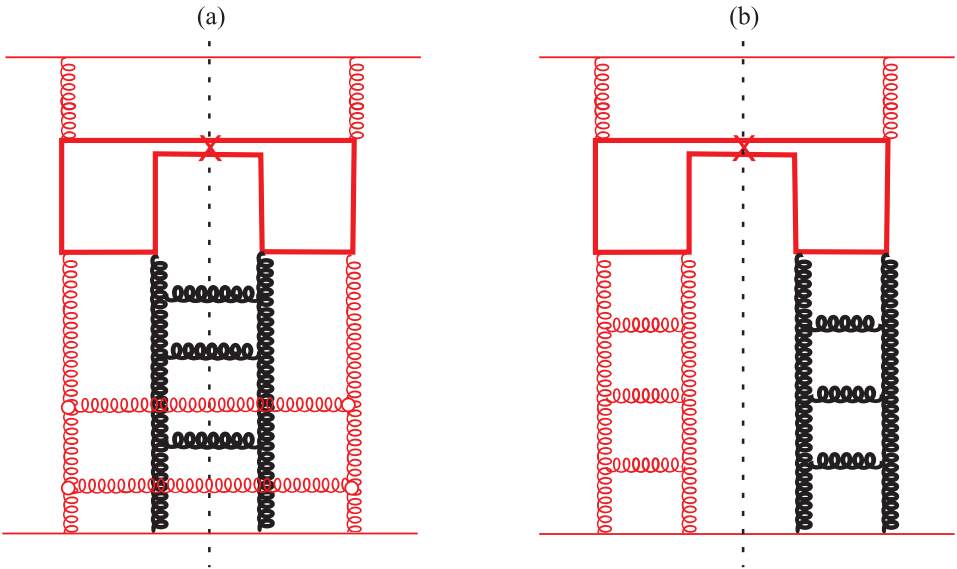}}
\caption{Prompt $J/\psi$ (a) inelastic and (b) single diffractive production arising from an
interaction in which the two $t$-channel gluons belong to different Pomerons.
  One Pomeron is shown in bold, the other in fainter print.}
\label{fig:oct}
\end{center}
\end{figure}

 Besides the diagrams shown in Fig.~\ref{fig:S}, where two $t$-channel gluons
 in a symmetric colour octet state are placed rather close to each other
 in the impact parameter ($b_t$) plane, there may also be a contribution from
 diagrams like those shown in Fig.~\ref{fig:oct}. This may be viewed as the
 production of a colour-octet $c\bar c$ pair that subsequently changes
 colour via  rescattering. Here the two $t$-channel gluons in the
 amplitude belong to two different Pomerons, that is to two
 different parton showers. A convenient way to
 calculate such a contribution is to use the AGK cutting
 rules \cite{AGK}, that is to calculate diagram Fig.~\ref{fig:oct}b and
 then use the relation
 \begin{equation}
 \sigma^{{\rm Fig}.\ref{fig:oct}a}~=~2~\sigma^{{\rm Fig}.\ref{fig:oct}b}\; .
 \label{agk}
 \end{equation}
Strictly speaking, the two gluons in the lower (left or right) parts of
Fig.~\ref{fig:oct}b cannot form a colour-singlet (Pomeron) state.
 On the other hand, these two gluons are in a colour \emph{symmetric}
 state, and therefore we may use the AGK relation (\ref{agk}) to
 simplify the calculation.

The amplitude corresponding to Fig.~\ref{fig:oct}b is very similar to that of Fig.~\ref{fig:S}.
 The main difference is that now the density of the second
 $t$-channel gluon is given by an independent gluon distribution $xg(x)$.
 That is, we need the probability to find two gluons 
%$w_{gg}=(xg(x))^2$. 
$w_{gg} \approx (xg(x))^2$.
On the other hand these two gluons are uniformly distributed over the whole
 transverse area occupied by a proton. Therefore the integration over
 the transverse momentum $Q_t$ is limited by the `elastic' slope $B$,
 which may be taken from the slope observed for diffractive $J/\psi$
 photoproduction; that is $B\sim 4.5$ GeV$^{-2}$ \cite{slope}.
Thus we can consider the lowest-order amplitudes given by Eqs.~(\ref{Aa},\ref{Ab})
 at $Q_t=0$, and take the integral $\int dQ^2_t=1/B$. 

The factor $(N_c\alpha_s/\pi)^2$,
 together with two logarithmic integrations, $dl^2_t/l^2_t$,
in each amplitude\footnote{The lowest-order gluon distribution at low $x$ is given by
the first iteration of the DGLAP equation, that is
$xg(x)=(N_c\alpha_s/\pi)\int dl^2/l^2$}, becomes
 the gluon distribution $w_{gg}=(xg(x))^2$ in the cross section.
Next we have to account for an extra 1/2 in the colour factor, which is
 cancelled by the factor 2 in  the AGK relation (\ref{agk}).
  Thus, finally, we obtain
 \begin{equation}
 \frac{d\sigma^{{\rm Fig}.\ref{fig:oct}a}}{dy}\;=\;\frac{10\pi^4\alpha_s^3g^2}{3BM^6}
x_1g(x_1)x_2g(x_2)[x_1g(x_1)+x_2g(x_2)].
 \label{s4a}
 \end{equation}
Using MRST2001 LO gluons \cite{MRST}, we find that\footnote{We take
the scale $\mu=M/2$ both in the 
 gluon distribution and in the QCD coupling $\alpha_s$.}
 $\sigma^{{\rm Fig}.\ref{fig:oct}a}(|y|<0.6)=2.2~\mu$b, at the Tevatron
 energy $\sqrt S=1.96$ TeV.
  This is smaller than
the main contribution discussed in Section 3. However at larger energies
 the `two Pomeron' cross section will grow faster than the contribution of
 Section 3. Indeed
\begin{equation}
 \sigma^{{\rm Fig}.\ref{fig:oct}a}~\propto ~x_1g(x_1)(x_2g(x_2))^2~\propto ~
  x_1^{-\lambda}x_2^{-2\lambda}~\propto S^{3\lambda/2}
 \label{sigP}
 \end{equation}
 in the central region of small $y$, where $x_i\propto 1/\sqrt S$,
whereas the cross section (\ref{sigf}) is proportional
 to $S^{\lambda}\ln S$ only. In particular, at the LHC energy $\sqrt
 S=14$ TeV we expect in the centre of the rapidity plateau
 $d\sigma^{{\rm Fig}.\ref{fig:oct}a}(y=0)/dy=6.7~\mu$b, while the cross section (\ref{sigf}) is
 $d\sigma^{{\rm Fig}.\ref{fig:S}}(y=0)/dy=8.1~\mu$b.

At first sight, the transverse momentum distribution of the $J/\psi$ for the contributions
of Fig.~\ref{fig:oct} should be peaked at low $Q_t$, namely $Q^2_t\sim 1/B$.
This is true for the contribution of the diagram Fig.~\ref{fig:oct}b, but not
for the inelastic process, Fig.~\ref{fig:oct}a.  The emission of intermediate
gluons spreads out the $J/\psi$ distribution from the inelastic process, so
that we expect a spectra
\begin{equation}
\label{6Qt}
d\hat\sigma^{{\rm Fig}.\ref{fig:oct}a}/dQ^2_t~~\propto~~ g^2\alpha_s(M_\perp)^3 /M^6_\perp,
\end{equation}
analogous to (\ref{Qt}).

We must of course take care of possible double counting.  Note that the amplitudes shown
in Fig.~\ref{fig:oct}a and Fig.~\ref{fig:S} are very similar, and the gluon
densities given by the global parton analyses do not distinguish the gluons
coming from one or more parton showers.  Therefore, to be conservative, in what follows
we consider only the contribution of Fig.~\ref{fig:S}.  A possible contribution of Fig.~\ref{fig:oct}a
is well within the uncertainties of the lowest-order in  $\alpha_s$ calculations.

%Nevertheless it would be interesting to observe the single diffractive $J/\psi$ production
%of Fig.~\ref{fig:oct}b.  This process differs qualitatively from the other contributions.
%The predicted cross section $d\sigma(y=0)/dy$ is about
%0.55 pb for the single diffraction of one proton at the Tevatron.  This
%must be multiplied by the rapidity gap survival probability ${\hat S}^2 \simeq 0.15$ \cite{KMRsoft},
%which gives
%\be
%\left. {d \sigma^{{\rm SD}}  \over dy}\right\vert_{y=0}~\simeq ~ 80~{\rm nb}.
%\ee
%Unfortunately due to the peaked transverse momentum distribution, $Q_t\sim 1/\sqrt{B} \lesim 0.5$ GeV,
%the $J/\psi$ mesons in the {\it Pomeron fragmentation} region will have
%large pseudorapidity, $\eta \sim 5$, outside the acceptance of the central detector.
%Therefore it will not be easy to observe this single diffractive signal.

\subsection{Associative ($J/\psi\;+\; c\bar c$) production}

\begin{figure}
\begin{center}
\centerline{\epsfxsize=0.5\textwidth\epsfbox{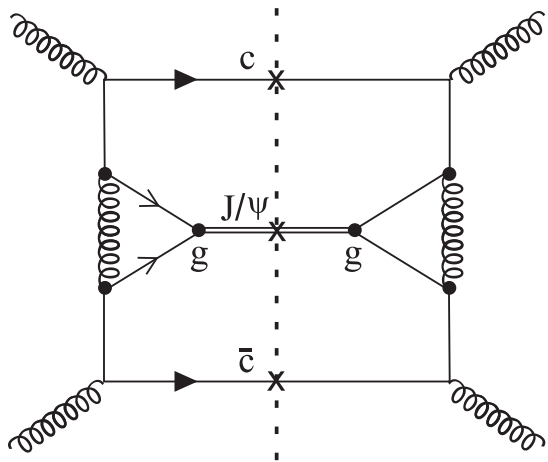}}
\caption{The diagram used to compute the associated production of $J/\psi$
together with a $c\bar c$ pair.  The $t$-channel gluon is needed in order
to put the virtual $t$-channel charm quarks on mass shell.}
\label{fig:cc}
\end{center}
\end{figure}

Another possibility is to consider the production of  a $J/\psi$ meson
 together with a $c\bar c$ pair, as shown in Fig.~\ref{fig:cc}.
 In this case the expected hard cross section at the Tevatron is about
 $\hat\sigma(c\psi\bar c)\sim 2$ nb, leading to
\begin{equation}
\left.  \frac{d\sigma(c\psi\bar c)}{dy}\right\vert_{y=0} ~=~\hat\sigma ~x_1g(x_1)~x_2g(x_2)~
 \sim ~0.05~\mu{\mbox b}\; .
 \label{cJc}
 \end{equation}
For inelastic hadroproduction, the contribution coming from
  this mechanism is
not large, of the order of 1\% . However in $e^+e^-$ annihilation,
 associative production is much more important, see for example Ref.~\cite{ABK}.
Indeed, in $e^+e^-$ annihilation there is no suppression
for the production of the first $c\bar c$-pair. On the other hand
 for time-like
 annihilation kinematics the third $t$-channel gluon ($l$ in Fig.~\ref{fig:H})
 can couple to the nearest parton only\footnote{Otherwise we destroy the
 leading logarithms.}. Thus we loose the enhancement caused by a
 large parton
 multiplicity, that is the factor $\Delta y\sim \ln(S/M^2)$ in
 (\ref{sigf}). Without this factor the contributions of
 single (Fig.~\ref{fig:H}) and associative (Fig.~\ref{fig:cc}) production are expected to
 be of the same order (in agreement with the
measurements of the Belle collaboration for $e^+e^-$ annihilation \cite{eePsi}).

Note that if the inelastic $J/\psi$ hadroproduction were to originate
from `colour-octet' dynamics, that is the main yield of $J/\psi$
were to come from gluon $g^*\to J/\psi +...$
fragmentation \cite{Oct,Sal}, then we would expect the same ratio of `direct'
to `associative' contributions in $e^+e^-$ annihilation, as that for hadroproduction, 
contrary to our perturbative QCD predictions.

\subsection{Production via $\chi_c$ and $b\bar b$ decays}

Finally we have the possibility of non-prompt $J/\psi$ production.
Experimentally it is observed that a fraction of $J/\psi$ mesons originate from
$\chi \to J/\psi+\gamma$ and $b \to J/\psi+X$ decays.  We do not discuss here
the details of these production mechanisms.  However in order to predict
the {\it total} $J/\psi$ yield we estimate these contributions using the
experimental data of Refs.~\cite{chi,bb} and collinear factorization.
Symbolically, the relation that we use is of the form\footnote{The relation
also happens to hold for $\chi_1$ production, despite the fact that it cannot
be formed by gluon-gluon fusion.  The dominant mechanism for $\chi_1$
production is the process with antisymmetric colour-octet $gg$ exchange.
However, since this exchange corresponds to gluon Reggeization, it has
the same energy (and $x$) dependence as the exchange of a single gluon.}
\be
d\sigma/dy~=~x_1 g(x_1)~{\hat \sigma}~x_2 g(x_2),
\label{eq:fac}
\ee
where $y=\frac{1}{2}{\rm ln}(x_1/x_2)$ and the normalisation of $\hat\sigma$ is adjusted
to fit the $\chi$ and $b\bar b$ data.
Since the energy dependence of the cross section is driven simply by the $x$ dependence
of the gluon densities, the experimental data at one fixed energy are
sufficient to estimate these non-prompt contributions at other
energies\footnote{In Ref.~\cite{chi} the cross section was given for $\chi$ production
in the forward hemisphere.  Therefore to normalize the $J/\psi$ cross section,
we use (\ref{eq:fac}) to calculate the integrated cross section.}.

\section{Energy and rapidity dependence}
\begin{figure}
\begin{center}
\centerline{\epsfxsize=0.9\textwidth\epsfbox{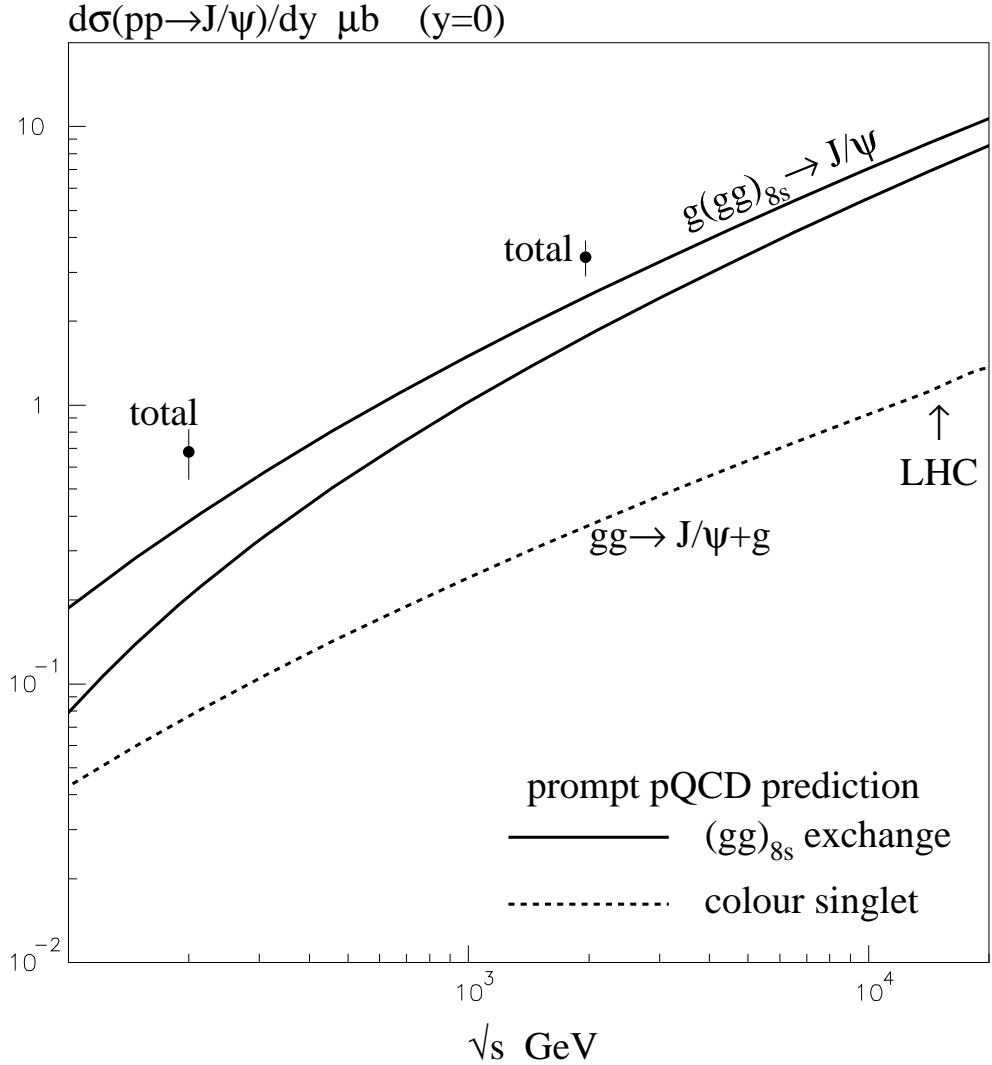}}
\caption{The energy dependence of {\it prompt} $J/\psi$ production obtained from the
colour-singlet mechanism, $gg \to J/\psi~g$, is shown by the dashed curve.
The contribution obtained from our subprocess, $g(gg)_{8s} \to J/\psi$, is shown by continuous curves with
$x_{\rm max}$ = 0.1 (lower) and 0.5 (upper).  Also shown are the
values for the {\it total} $J/\psi$ yield measured at RHIC \cite{RHIC} and the Tevatron \cite{CDF}.}
\label{fig:E1}
\end{center}
\end{figure}

In comparison with the usual perturbative QCD colour-singlet (Fig.~\ref{fig:PQCD}a)
and colour-evaporation mechanisms, where the
energy dependence is essentially determined by the product of the gluon densities, see
(\ref{eq:fac}), our perturbative contribution to the cross section (Fig.~\ref{fig:PQCD}b) is enhanced by
an additional log$S$ factor, $\Delta y$ in (\ref{sigf}).  Therefore the energy
dependence of these two contributions is different, as can be seen in Fig.~\ref{fig:E1}.
To demonstrate the uncertainty arising from the choice of the cut-off
$x_{\rm max}$ in the additional logarithm $\Delta y$ in the computation of the new perturbative
contribution,
we show predictions for the two values, $x_{\rm max}$ = 0.1 (lower curve) and 0.5 (upper curve).

Note that the amplitude of the new subprocess is calculated at lowest order in $\alpha_s$.  
It is not unusual to have a next-to-leading order $K$ factor
of the order of 2 for the production of relatively low-mass states, as in Drell-Yan production for example.
 On the other hand, we see no reason for a strong energy
dependence of such a $K$ factor.  So, noting the good agreement with the Tevatron data,
we expect our predictions at the LHC energy to be reliable.

\begin{figure}
\begin{center}
\centerline{\epsfxsize=0.9\textwidth\epsfbox{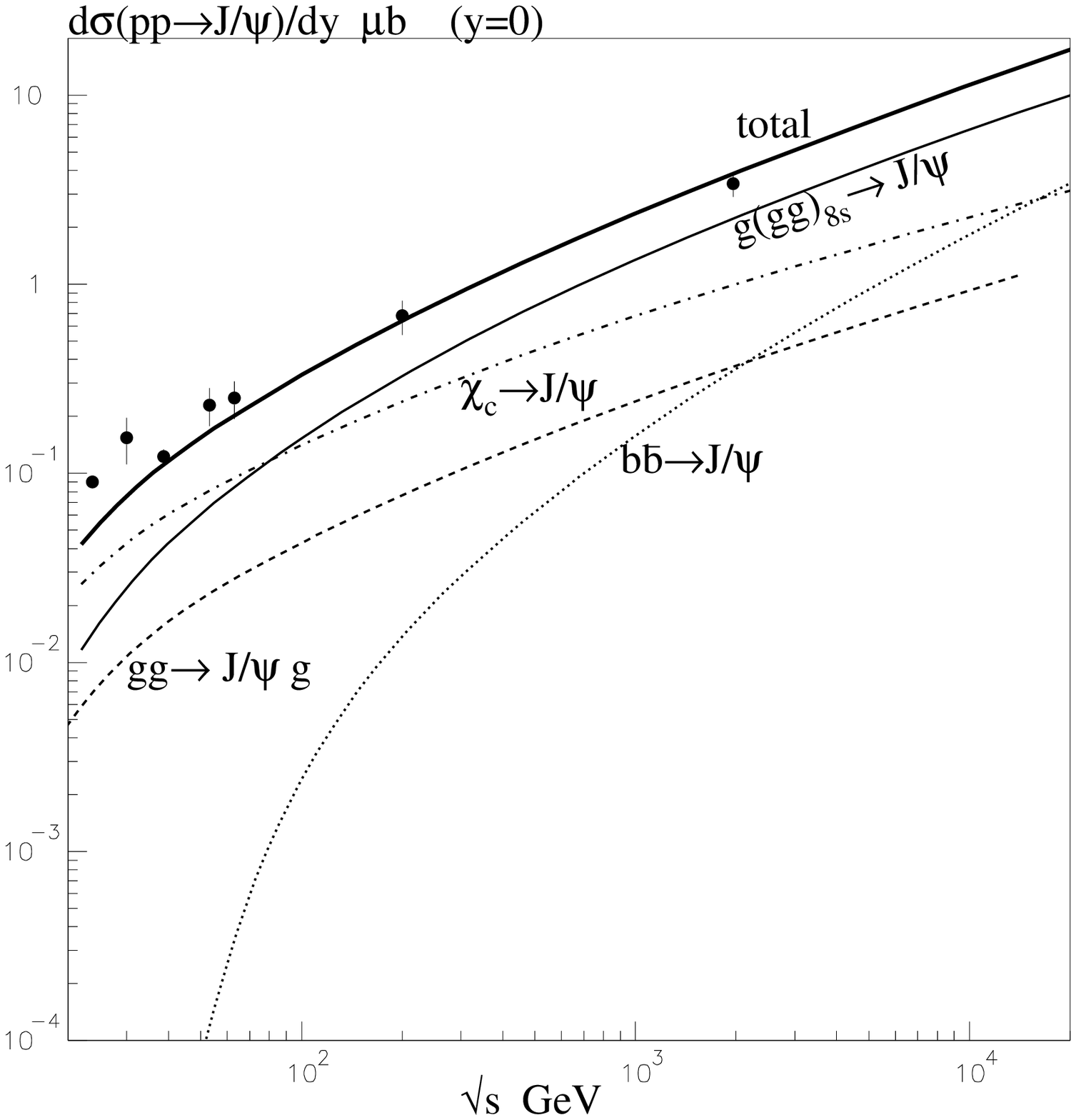}}
\caption{The energy dependence of inelastic $J/\psi$ production compared with a selection of the
available data from fixed-target $pN$ to Tevatron $p \bar p$ collider energies 
\cite{exp,RHIC,CDF}. Besides the total (bold curve), we also display the prompt ($g(gg)_{8s} \to J/\psi$
and $gg \to J/\psi~g$) and non-prompt ($\chi_c \to J/\psi$ and $b\bar b \to J/\psi$) components.}
\label{fig:E2}
\end{center}
\end{figure}

The energy behaviour of the total inelastic $J/\psi$ cross section in the fixed-target/ISR to Tevatron energy interval
is shown in Fig.~\ref{fig:E2}, together with the components from the following
subprocesses: $g(gg)_{8s} \to J/\psi,~~
gg \to J/\psi~g,~~ \chi_c \to J/\psi$ and $b\bar b \to J/\psi$.
We see that the cross section is slightly underestimated at the lower energies, where a contribution
initiated by the $q\bar q$ subprocess, that is by secondary Reggeons, may have some
influence.  Indeed, it was
noted in Ref.~\cite{MRS} that the difference between the $J/\psi$ production cross sections from $pp$ and
$p\bar p$ interactions indicates a noticeable $q\bar q$ contribution at the lower energies.

\begin{figure}
\begin{center}
\centerline{\epsfxsize=0.9\textwidth\epsfbox{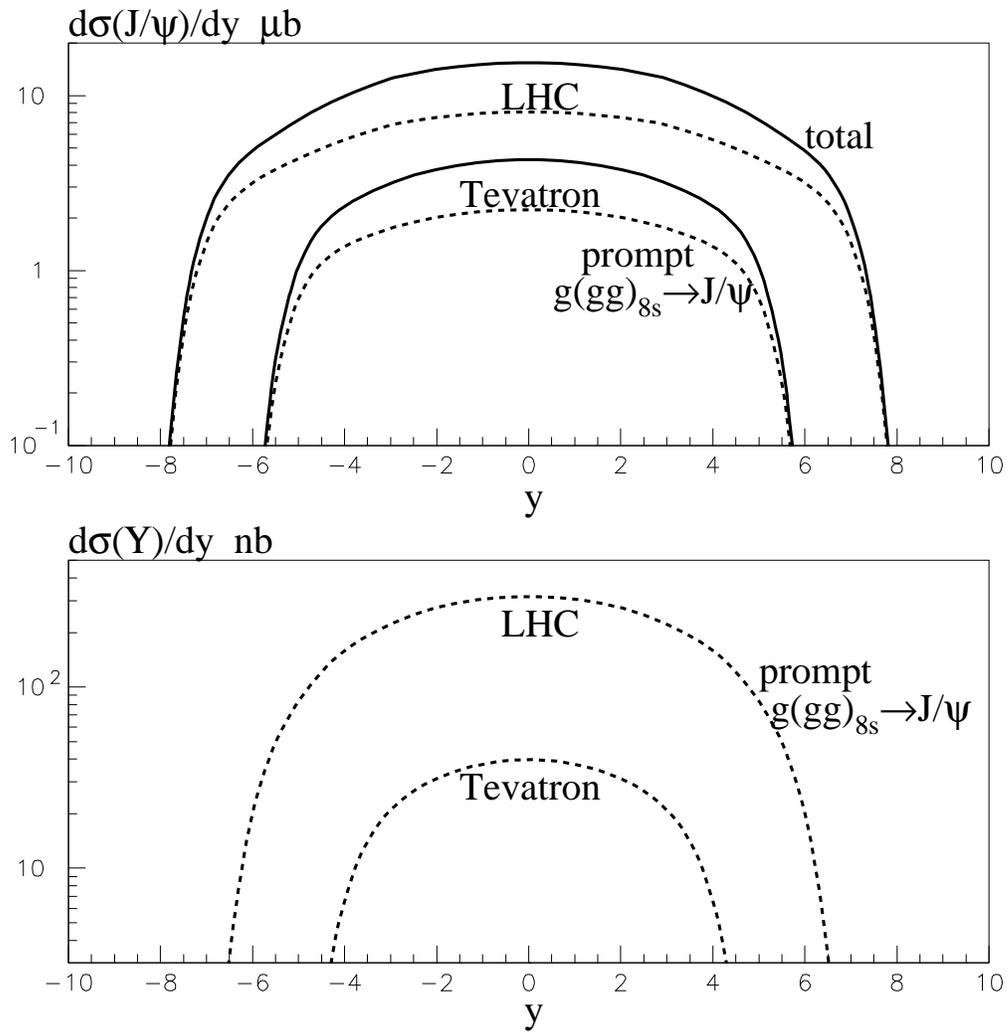}}
\caption{The rapidity distributions of $J/\psi$ and  $\Upsilon(1S)$ production at Tevatron
and LHC energies.   The continuous and dashed curves correspond to the total and prompt
yields respectively.}
\label{fig:y}
\end{center}
\end{figure}

In Fig.~\ref{fig:y} we show predictions for the rapidity distributions of $J/\psi$ and $\Upsilon(1S)$
production at both the Tevatron (1.96 TeV) and LHC\footnote{Note that the General Purpose Detectors ATLAS and CMS can only measure the high transverse momentum tail of quarkonium production, the majority of the final-state
leptons falling below the trigger thresholds. The ALICE detector, on the other hand, is ideally suited to a measurement of both charm and bottom quarkonium cross sections~\cite{Vogt,ALICE}.  ALICE can measure electrons and muons
down to very low transverse momentum ($ {\cal O}(1 \; {\rm GeV}/c)$) in the pseudorapidity ranges
$|\eta_e| < 0.9$ and $2.5 < |\eta_\mu| < 4.0$ respectively. This gives non-zero acceptance for the 
$J/\psi$ down to $p_T(J/\psi) = 0$, see Figs.~60 and 63 of Ref.~\cite{Vogt}.}  (14 TeV) energies.
In Table~\ref{table1}  we compare our results for the total $J/\psi$ and $\Upsilon(1S)$ cross sections at LHC
with the predictions of the CEM (taken from Ref.~\cite{Vogt}).
Note that our predicted
cross sections are systematically larger because of the steeper $\sqrt{S}$ dependence
caused by the $\log(S/M^2_\perp)$ factor in Eq.~(\ref{sigf}).
\begin{table}$$\begin{array}{|l @{\qquad}|  c | c |} \hline
 &  \mbox{pQCD} & \mbox{CEM}  \rule[-1.5ex]{0ex}{5ex}  \\ \hline
 \sigma_{J/\psi}\mbox{(prompt)} \ \ (\mu{\rm b}) & 80 &  33  \\
 \sigma_{J/\psi}\mbox{(total)}  \ \ (\mu{\rm b})&  150 & 54 \\
 \sigma_{\Upsilon(1S)}\mbox{(prompt)} \ \ (\mu{\rm b}) & 2.5 & 0.4 \\\hline
\end{array}$$
\caption{The predictions for $J/\psi$ and $\Upsilon(1S)$ cross sections (in $\mu$b) at the LHC. 
For the $J/\psi$, the `total' cross section includes the additional contributions from 
$\chi_c$ and $b\bar b$ decay. The CEM predictions are taken from Tables 9 and 10 in Ref.~\cite{Vogt}.}
\label{table1}
\end{table}

\section{Conclusions}

We have calculated the prompt hadroproduction of $J/\psi,~ \psi'$ and
$\Upsilon(1S,2S,3S)$ states within a perturbative QCD framework, without
any non-perturbative contributions (such as occur explicitly in the NRQCD colour-octet model
and implicitly in the CEM).
Recall that the original colour-singlet LO perturbative contribution (based on
the subprocess $gg \to J/\psi~g$) falls
well short of the data.  However, here, we have studied another perturbative QCD
contribution, which turns out to be dominant.  The basic subprocess is
$g(gg)_{8s} \to J/\psi$, see Figs.~\ref{fig:H},\ref{fig:S} or Fig.~\ref{fig:PQCD}b.  This contribution is enhanced,
particularly at high energies, since the additional $t$-channel gluon can
couple to a large number of parton spectators.

The uncertainties of such a computation are listed at the end of Section 3.
They are not small.  However, with our natural choices of scale and of the effective
gluon mass, we successfully describe the available high-energy RHIC and Tevatron $J/\psi$ data.
In addition, without any new parameters, we obtain an excellent description of the $\psi'$ data,
and even a satisfactory description of $\Upsilon$ production.

There is additional qualitative support for the $g(gg)_{8s} \to J/\psi$ mechanism
coming from the measurement of the $J/\psi$ polarization at the Tevatron.  This mechanism predicts a
longitudinal polarization at large $Q_t$, in agreement with the data, whereas the
colour-octet model leads to a transverse polarization \cite{polthy}.

\section*{Acknowledgements}

We thank Anatoli Likhoded, Lev Lipatov and Orlando Villalobos-Baillie for useful discussions.
ADM thanks the Leverhulme Trust for an Emeritus Fellowship and MGR thanks the IPPP at the University of
Durham for hospitality. This work was supported by
the UK Particle Physics and Astronomy Research Council, by a Royal Society special
project grant with the FSU, by grant RFBR 04-02-16073
and by the Federal Program of the Russian Ministry of Industry, Science and Technology
SS-1124.2003.2.

\end{document}